\documentclass[aps,prl,twocolumn,showpacs]{revtex4}
\usepackage{graphicx}
\usepackage{dcolumn}

\begin{document}

\preprint{APS/123-QED}

\title{Magneto-structural properties of the layered quasi-2D triangular-lattice antiferromagnets Cs$_2$CuCl$_{4-x}$Br$_x$ for \textit{x} = 0,1,2 and 4}
\author{S. K. Thallapaka$^{1}$}
\author{B. Wolf$^{1}$}
\author{E. Gati$^{1}$}
\author{L. Postulka$^{1}$}
\author{U. Tutsch$^1$}
\author{B. Schmidt$^{2}$}
\author{P. Thalmeier$^2$}
\author{F. Ritter$^{1}$}
\author{C. Krellner$^{1}$}
\author{Y. Li$^{3}$}
\author{V. Borisov$^{3}$}
\author{R. Valent\'{\i}$^3$}
\author{M. Lang$^{1}$}

\address{$^{1}$Physikalisches Institut, J.W. Goethe-Universit\"{a}t Frankfurt(M), SFB/TR 49, D-60438 Frankfurt(M), Germany}
\address{$^{2}$Max-Planck-Institut f\"{u}r Chemische Physik fester Stoffe, D-01187 Dresden, Germany}
\address{$^{3}$Institut f\"ur Theoretische Physik, J.W. Goethe-Universit\"{a}t Frankfurt(M), SFB/TR 49, D-60438 Frankfurt(M), Germany}


\begin{abstract}
 We present a study of the magnetic susceptibility $\chi_{mol}$ under variable hydrostatic pressure on single crystals of Cs$_2$CuCl$_{4-x}$Br$_x$. This includes the border compounds \textit{x} = 0 and 4, known as good realizations of the distorted triangular-lattice spin-1/2 Heisenberg antiferromagnet, as well as the  isostructural stoichiometric systems Cs$_2$CuCl$_{3}$Br$_1$ and Cs$_2$CuCl$_{2}$Br$_2$. For the determination of the exchange coupling constants $J$ and $J^{\prime}$, $\chi_{mol}$ data were fitted by a $J-J^{\prime}$ model \cite{Schmidt2015}. Its application, validated for the border compounds, yields a degree of frustration $J^{\prime}$/$J$ = 0.47 for Cs$_2$CuCl$_3$Br$_1$ and $J^{\prime}$/$J$ $\simeq$ 0.63 - 0.78 for Cs$_2$CuCl$_2$Br$_2$, making these systems particular interesting representatives of this family. From the evolution of the magnetic susceptibility under pressure up to about 0.4\,GPa, the maximum pressure applied, two observations were made for all the compounds investigated here. First, we find that the overall energy scale, given by $J_c = (J^2$ + $J^{\prime 2}$)$^{1/2}$, increases under pressure, whereas the ratio $J^{\prime}$/$J$ remains unchanged in this pressure range. These experimental observations are in accordance with the results of DFT calculations performed for these materials. Secondly, for the magnetoelastic coupling constants, extraordinarily small values are obtained. We assign these observations to a structural peculiarity of this class of materials.
\end{abstract}


\maketitle

\section{Introduction}
The Cu-halide systems Cs$_2$CuCl$_4$ and Cs$_2$CuBr$_4$ have been established as good realizations \cite{Coldea2001,Ono2003} of a layered anisotropic triangular-lattice spin-1/2 Heisenberg antiferromagnet with the exchange constants $J$ and $J^{\prime}$ (see inset Fig.1), a paradigmatic model in quantum magnetism. Both systems crystallize in an orthorhombic structure, space group $Pnma$, where Cu$^{2+}$ (\textit{S} = 1/2) ions form a distorted triangular lattice parallel to the $bc$ plane. These systems have been of ongoing interest because of their intriguing properties resulting from the interplay of strong quantum fluctuations and geometrical frustration, quantified by the ratio of the exchange coupling constants $J^{\prime}/J$. Despite their structural similarities, the systems show remarkably different magnetic behavior which has been attributed to their different degree of frustration. According to high-field studies of inelastic neutron scattering and ESR, the $J^{\prime}/J$ ratio amounts to $\simeq$ 0.34 for Cs$_2$CuCl$_4$\cite{Coldea2002} and $\simeq$ 0.41 for Cs$_2$CuBr$_4$ \cite{Zvyagin2014}. Cs$_2$CuBr$_4$ undergoes a transition into a helical long-range-ordered antiferromagnetic state below $T_N$ = 1.4\,K \cite{Ono2003}. Upon applying an external magnetic field, a plateau is found in the magnetization at about one third of the saturation magnetization \cite{Ono2003} before the fully polarized state is reached around 30\,T. In contrast, for Cs$_2$CuCl$_4$, the long-range order occurs at $T_N$ = 0.62\,K and is preceded by a range characterized by spin-liquid behavior \cite{Coldea2001,Coldea2003}. Moreover, indications for a quantum-critical point near 8.5\,T ($B\|a$) were observed, the critical properties of which are consistent with Bose-Einstein condensation of triplons \cite{Coldea2002,Radu2005}. Motivated by these observations, ultrasonic investigations were performed to study the interplay between spin-lattice interactions and quantum criticality \cite{Cong2016}. It was found that the elastic data at temperatures below 0.1\,K could be surprisingly well described by a classical phenomenological model, indicating the presence of small disturbing interactions which drive the system away from quantum criticality, and a very small magnetoelastic coupling constant \cite{Cong2016}.

In an attempt to modify the magnetic properties via chemical substitution, compounds of the Cs$_2$CuCl$_{4-x}$Br$_x$ series for 0 $\leq$ \textit{x} $\leq$ 4 were synthesized \cite{Kruger2010}. Subsequent investigations showed that there is a striking discontinuous evolution of the magnetic \cite{Cong2011} and structural \cite{Kruger2010,Well2015} properties upon increasing the Br content, which was assigned to a site-selective substitution of the halide sublattice \cite{Cong2011}. Importantly, these investigations suggested that Cs$_2$CuCl$_3$Br$_1$ and Cs$_2$CuCl$_2$Br$_2$ mark particularly interesting new compounds, characterized by a well-ordered halide sublattice, i.e., a regular local Cu environment, and a larger degree of frustration compared to the pure chlorine (\textit{x} = 0) material.

The aim of the present work is to provide a basic magneto-structural characterization of the stoichiometric members of the Cs$_2$CuCl$_{4-x}$Br$_x$ series. In our study, which also includes the border compounds \textit{x} = 0 and 4, we especially focus on the recently discovered systems Cs$_2$CuCl$_3$Br$_1$ and Cs$_2$CuCl$_2$Br$_2$. All members of this family share Jahn-Teller distorted (CuCl$_4$)$^{2-}$ tetrahedra as the relevant structural unit, which are well isolated from each other by not sharing any common coordination element. As an experimental tool, we use magnetic susceptibility measurements combined with hydrostatic (He-gas) pressure. The data will be analyzed by taking into account structural data under pressure, available for Cs$_2$CuCl$_4$ \cite{Xu2000}.

\section{Experimental}
Measurements of the magnetic susceptibility were performed by using a commercial superconducting quantum interference device (SQUID) magnetometer (Quantum Design MPMS). For the measurements under pressure up to 0.4\,GPa, a CuBe pressure cell was used which is connected to a room-temperature He-gas compressor. The compressor also serves as a gas reservoir to maintain $p$ $\approx$ const. conditions upon temperature sweeps. The dimensions of the CuBe cell are 8.6\,mm (outer diameter), 3.0\,mm (inner diameter) and 120\,mm length. An indium sample serves for an \textit{in situ} pressure determination \cite{Jennings1958}. The use of helium as a pressure-transmitting medium ensures truly hydrostatic pressure conditions in those $p$-$T$ ranges where helium is in its liquid phase. Even for temperatures below the solidification line, helium is still well-suited for pressure studies as it forms a comparatively soft solid as a consequence of its purely van-der-Waals-bonds \cite{Manna2012}. As described in detail in Ref. \cite{Kruger2010}, the Cs$_2$CuCl$_{4-x}$Br$_x$ (0 $\leq$ \textit{x} $\leq$ 4) crystals were grown from an  aqueous solution at 50$^\circ$C by using an evaporation method. The structural characterization revealed similar results as reported in Ref. \cite{Kruger2010,Well2015}.  The molar susceptibility $\chi_{mol}$ of all samples under investigation was taken in a field of 0.1\,T for $B\|b$-axis. The data have been corrected for a diamagnetic core contribution and the contribution of the pressure cell. Both contributions amount to less than 0.3 \% of the total signal in the temperature range under investigation. \\

\section{Pressure dependence of the magnetic susceptibility}
\begin{figure}
\includegraphics[width=0.95\columnwidth]{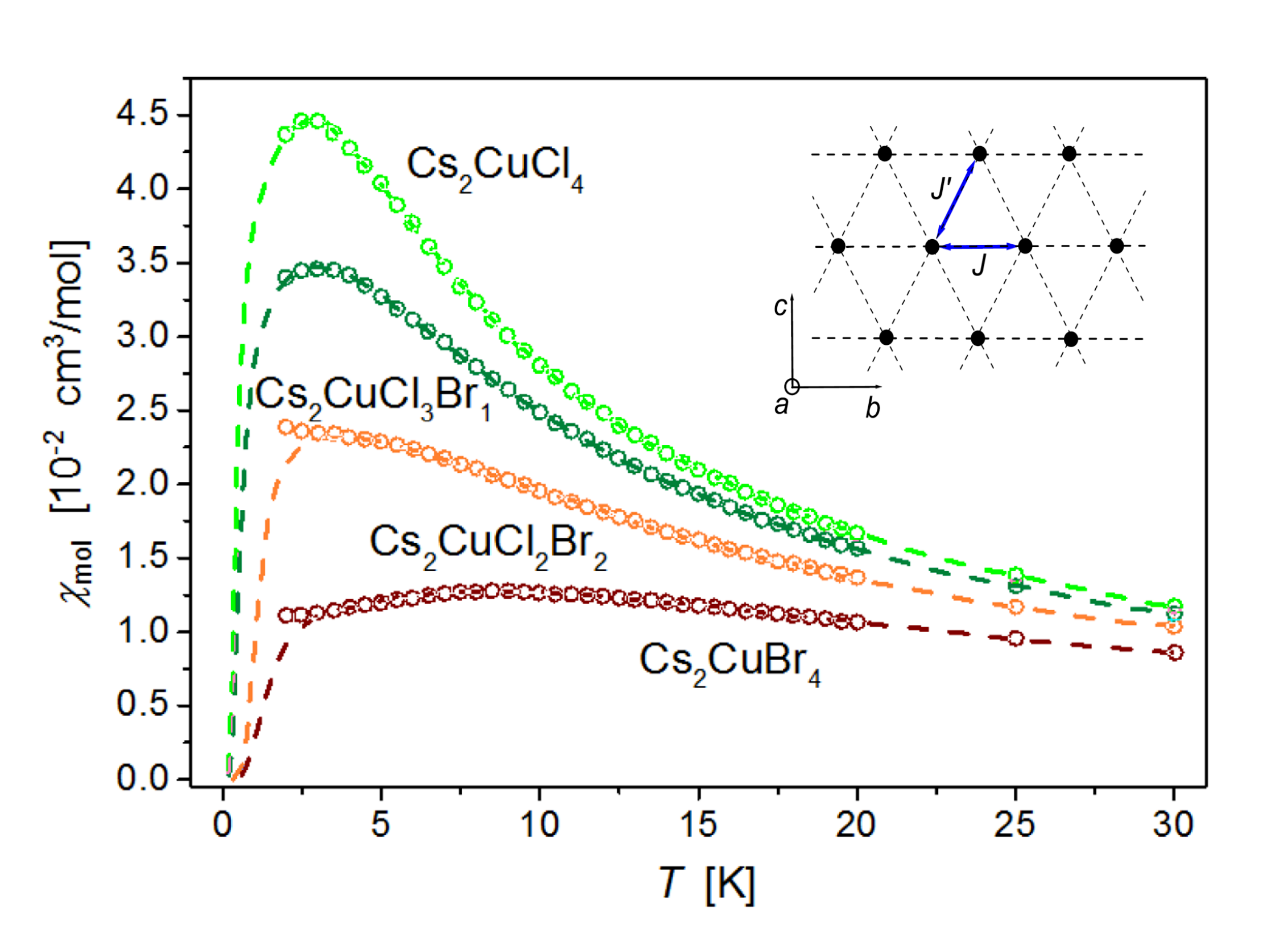}
\caption{(\emph{Color online}) Molar susceptibility $\chi_{mol}$  of Cs$_2$CuCl$_4$ (light green), Cs$_2$CuCl$_3$Br$_1$ (dark green), Cs$_2$CuCl$_2$Br$_2$ (orange) and Cs$_2$CuBr$_4$ (brown) as a function of temperature for 2\,K $\leq T \leq$ 30\,K. The broken lines represent fits to the data according the theoretical model described in Ref.\,\cite{Schmidt2015} by using the same color code. The inset shows the magnetic model, a distorted triangular lattice with exchange coupling constants $J$ and $J^{\prime}$ in the \textit{bc} plane, used to model the magnetic behavior for the Cs$_2$CuCl$_{4-x}$Br$_x$ compounds.}
\label{fig:1a}
\end{figure}

 In Fig.\,\ref{fig:1a} we give an overview of the molar magnetic susceptibilities, $\chi_{mol}$, for the Cs$_2$CuCl$_4$, Cs$_2$CuCl$_3$Br$_1$, Cs$_2$CuCl$_2$Br$_2$ and Cs$_2$CuBr$_4$ single crystals investigated here. The data, taken at ambient pressure, are consistent with published results \cite{Cong2011,Tokiwa2006}. The main features include a Curie-Weiss-like increase of $\chi_{mol}$($T$) with decreasing temperature followed by a rounded maximum at $T_{max}$ = 2.8\,K, 2.95\,K and 8.75\,K for Cs$_2$CuCl$_4$, Cs$_2$CuCl$_3$Br$_1$, and Cs$_2$CuBr$_4$, respectively. For Cs$_2$CuCl$_2$Br$_2$, a somewhat different behavior is observed. Here $\chi_{mol}$($T$) keeps growing upon cooling so that only a shoulder-like anomaly is visible at around 5\,K. The reason for this upturn is unclear at present. We stress that a simple Curie contribution due to isolated spins can be ruled out as the upturn neither can be suppressed in a larger magnetic field of 5\,T nor does it show any sample-to-sample variation.

\begin{figure}
\includegraphics[width=0.95\columnwidth]{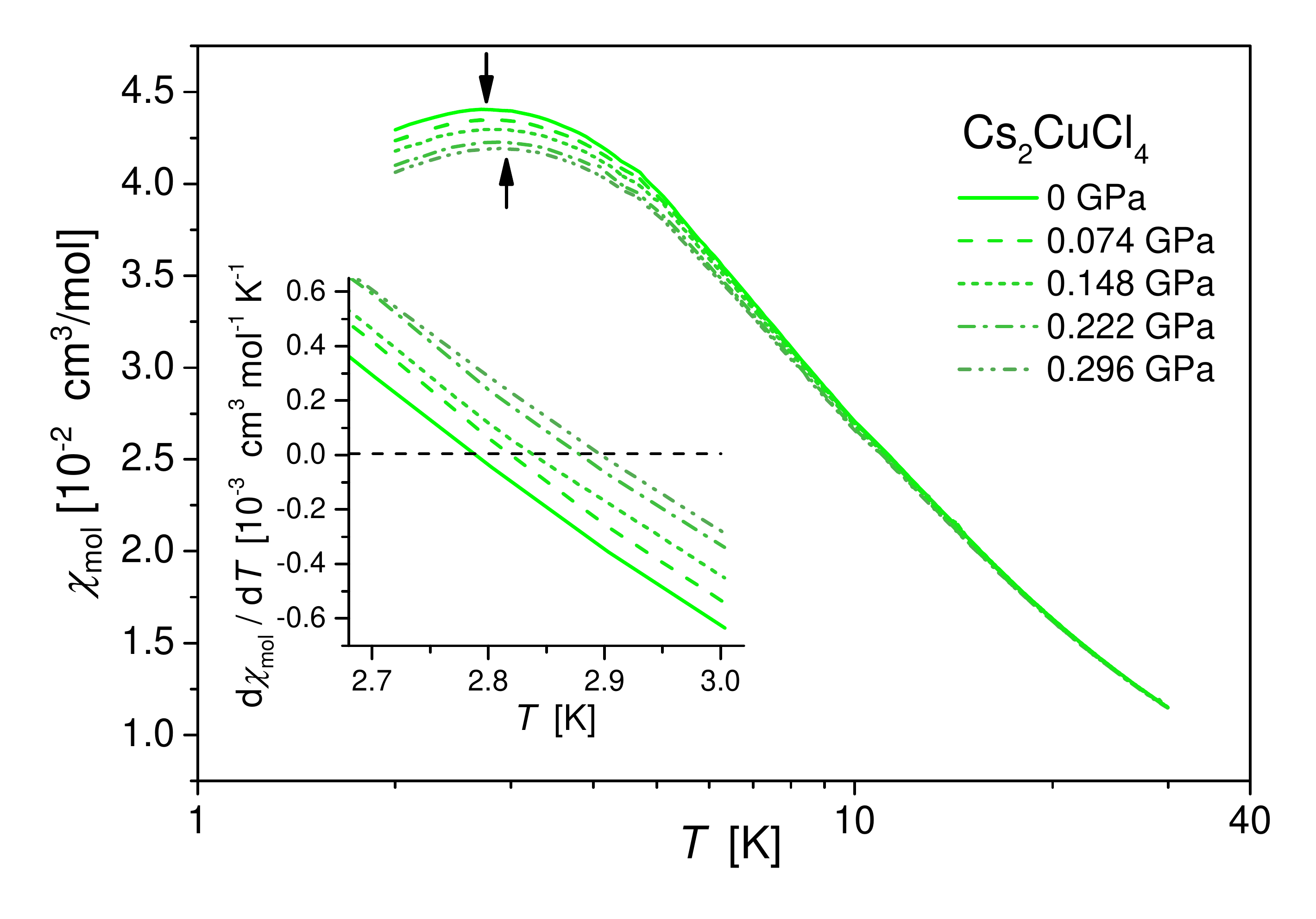}
\caption{(\emph{Color online}) Molar susceptibility $\chi_{mol}$ of Cs$_2$CuCl$_4$ as a function of temperature for 2\,K $\leq T \leq$ 30\,K under hydrostatic pressures up to 0.296\,GPa in a semi-logarithmic representation. The inset shows the same data as d$\chi_{mol}$/d\textit{T} for 2.7\,K $\leq T \leq$ 3.0\,K using the line types as in the main panel.}
\label{fig:1}
\end{figure}

In the following, we focus on the effect of hydrostatic pressure on the susceptibility. To this end, we show the low-temperature $\chi_{mol}$ data for 2\,K $\leq \textit{T} \leq$ 30\,K in a semi-logarithmic representation for the three different compounds at varying hydrostatic pressures. Starting with Cs$_2$CuCl$_4$ in Fig.\,\ref{fig:1}, we find a gradual suppression of $\chi_{mol}$($T$) with increasing pressure up to 0.296\,GPa, the maximum pressure applied here. As can be inferred from the inset of Fig.\,\ref{fig:1}, showing the derivative of the various data sets around the maximum position, d$\chi_{mol}$($T$)/d$T$, this suppression of the susceptibility is accompanied by a continuous shift of $T_{max}$ to higher temperatures at a rate d$T_{max}$/d$p$ = (0.38 $\pm$ 0.02)\,K/GPa.

\begin{table*}[t]
  \caption{Compilation of parameters which determine the magnetic behavior of the quasi-2D triangular-lattice Heisenberg antiferromagnets Cs$_2$CuCl$_{4-x}$Br$_x$ with \textit{x} = 0, 1, 2, 4 obtained from fits to magnetic susceptibility data according to Ref.\,\cite{Schmidt2015}. The magnetic coupling constants \textit{J} and \textit{J$^\prime$} are determined from the fit parameters \textit{J$_c$} and $\phi$. If not state otherwise, typical uncertenties are below 3\% for the $J$ values and less than 1 \% for the g values.}
\begin{tabular}{r|rcr|cccc}
\hline
  & \textit{J$_c$/$k_B$}\,(K) & \textbf{$\phi$}/$\pi$  & g (\textit{B}$\|$b) & \textit{J/$k_B$}\,(K)& \textit{J$^\prime$/$k_B$} \,(K) & & \textit{J$^\prime$}/\textit{J}  \\
\hline
Cs$_2$CuCl$_4$ & 4.75 & 0.40 & 2.09 & 4.52 & 1.47 & & 0.32 \\
Cs$_2$CuCl$_3$Br$_1$ & 6.03 & 0.36 & 2.11 & 5.46 & 2.57 & & 0.47 \\
Cs$_2$CuCl$_2$Br$_2$ & 8.81 & 0.32 - 0.29& 2.09 & 7.44 - 6.95& 4.72 -5.42 & & 0.63 - 0.78 \\
Cs$_2$CuBr$_4$ & 15.66  & 0.38 & 2.06 & 14.78  & 5.76 & & 0.40  \\
\hline
\end{tabular}
\end{table*}

\begin{figure}
\includegraphics[width=0.95\columnwidth]{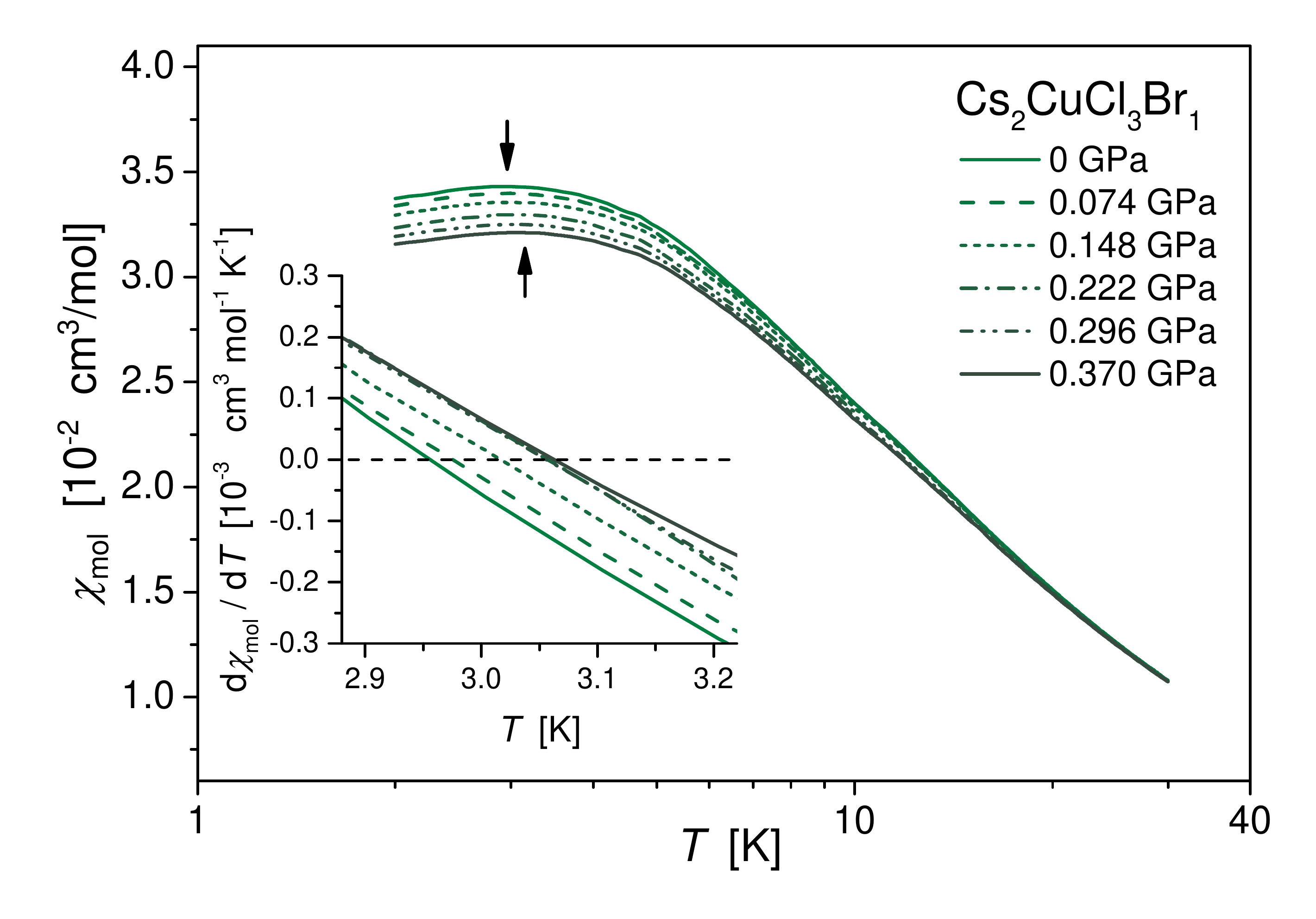}
\caption{(\emph{Color online}) Molar susceptibility $\chi_{mol}$ of Cs$_2$CuCl$_3$Br$_1$ as a function of temperature for 2\,K $\leq T \leq$ 30\,K under hydrostatic pressures up to 0.37\,GPa in a semi-logarithmic representation. The inset shows the same data as d$\chi_{mol}$/d\textit{T} for 2.9\,K $\leq T \leq$ 3.2\,K using the same line types as in the main panel.}
\label{fig:2}
\end{figure}

A qualitatively similar pressure effect is revealed for Cs$_2$CuCl$_3$Br$_1$, cf.\,Fig.\,\ref{fig:2}, where a maximum pressure of 0.37\,GPa was applied. As for Cs$_2$CuCl$_4$, the data reveal a gradual suppression of $\chi_{mol}$($T$) accompanied by a weak increase of $T_{max}$ at a rate d$T_{max}$/d$p$ = (0.29 $\pm$ 0.05)\,K/GPa.

\begin{figure}
\includegraphics[width=0.95\columnwidth]{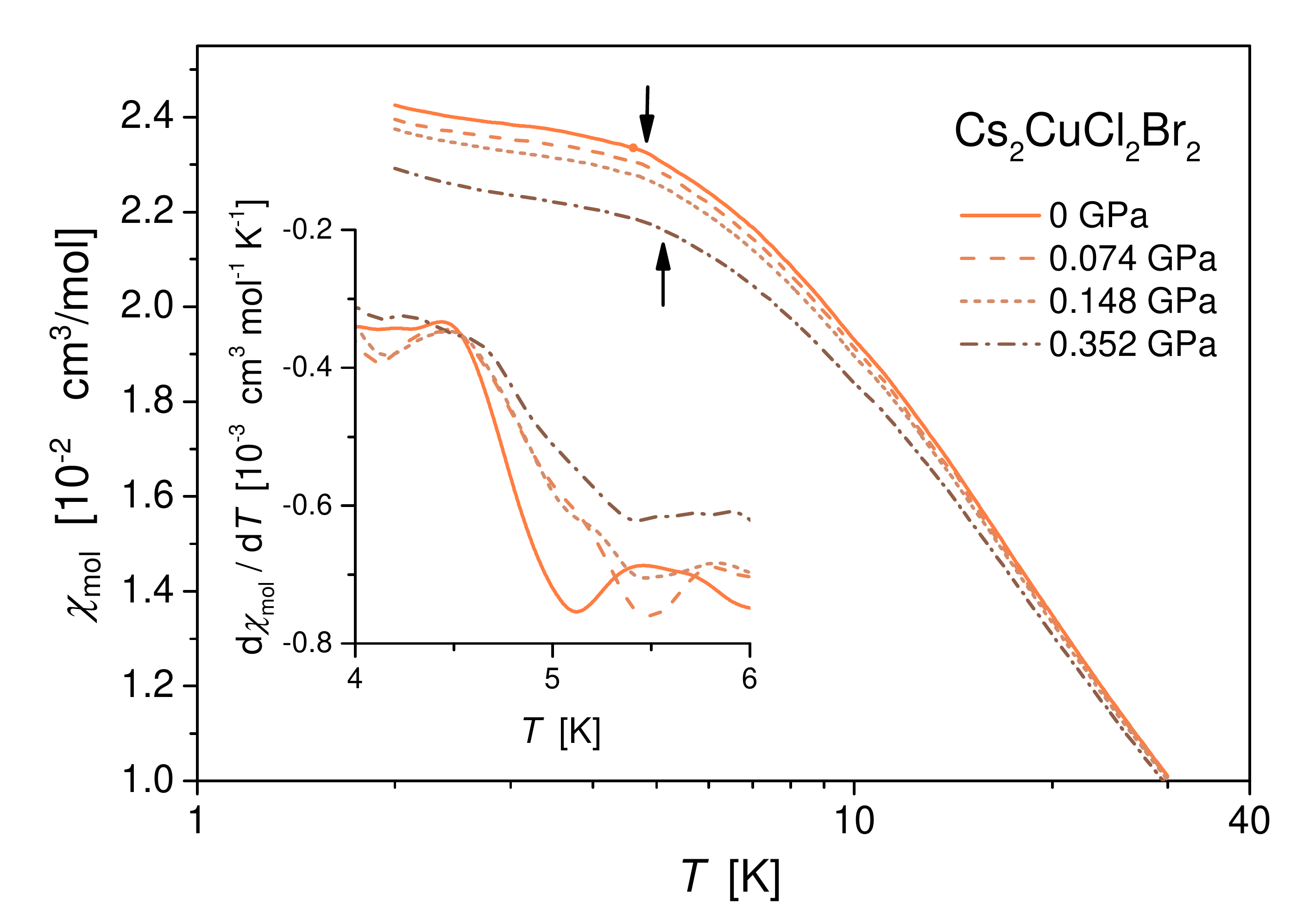}
\caption{(\emph{Color online}) Molar susceptibility $\chi_{mol}$ of Cs$_2$CuCl$_2$Br$_2$ as a function of temperature for 2\,K $\leq T \leq$ 30\,K under hydrostatic pressure up to 0.352\,GPa. The inset shows the same data as d$\chi_{mol}$/d\textit{T} for 4.0\,K $\leq T \leq$ 6.0\,K using the same line types as in the main panel.}
\label{fig:4}
\end{figure}

The corresponding $\chi_{mol}$(\textit{T}) data for Cs$_2$CuCl$_2$Br$_2$ are shown in Fig.\,\ref{fig:4} for pressures up 0.352\,GPa. Also for this composition, we find a weak suppression of $\chi_{mol}$($T$) with increasing pressure. In order to quantify the effect of pressure, we use the shoulder in $\chi_{mol}$ as the characteristic temperature, indicated by an arrow in the main panel -- a justification of this assignment will be given in the discussion part. We determine the position of the shoulder by evaluating the d$\chi_{mol}$($T$)/d$T$ data for the different pressures (inset of Fig.\,\ref{fig:4}) and applying the following criterion: as there is a step-like change in d$\chi_{mol}$($T$)/d$T$ between two regimes of almost constant slope, we use the midpoint of this jump as a measure of the position of the shoulder $T_s$. By applying this criterion, we find d$T_{s}$/d$p$ = (0.77 $\pm$ 0.18)\,K/GPa.

\begin{table*}[t]
  \caption{ The overall energy scale $J_c = (J^2$ + $J^{\prime 2}$)$^{1/2}$ and the $g$-factor both at $p_{max}$ together with the magnetoelastic coupling constant $k_{B}^{-1}$d$J_c$/d$\epsilon_V$ for Cs$_2$CuCl$_4$ and the stoichiometric compounds Cs$_2$CuCl$_3$Br$_1$ and Cs$_2$CuCl$_2$Br$_2$. If not state otherwise, typical uncertenties are below 3\% for the $J_c$ values and less than 1 \% for the g values. }
\begin{tabular}{r|rrrr}

\hline
& \textit{J$_c$/$k_B$}(\textit{p}$_{max}$) \,(K) & g(\textit{p}$_{max}$) (\textit{B}$\|$b) & $k_B^{-1}$d$J_c$/d$\epsilon_V$\,(K) \\
\hline
Cs$_2$CuCl$_4$ & 4.86 & 2.08 & 5.7 $\pm$ 0.4    \\
Cs$_2$CuCl$_3$Br$_1$ & 6.24 & 2.09 &  4.4 $\pm$ 0.8  \\
Cs$_2$CuCl$_2$Br$_2$ & 9.31 & 2.08 & 11.6 $\pm$ 2.8  \\
\hline
\end{tabular}
\end{table*}

\section{Modeling}
In Ref.\,\cite{Schmidt2015} it was shown that the magnetic susceptibility of the border compounds Cs$_2$CuCl$_4$ and Cs$_2$CuBr$_4$ can be very well described by a 2D $J-J^{\prime}$ triangular lattice $S$ = 1/2 Heisenberg antiferromagnet. According to this work, the low-temperature susceptibility with its characteristic maximum at $T_{max}$ and a height $\chi_{mol}^{max}$ = $\chi_{mol}$(\textit{T$_{max}$}), is determined by an overall energy scale $J_c = (J^2$ + $J^{\prime 2}$)$^{1/2}$, and the anisotropy parameter $\phi$ = tan$^{-1}$($J$/$J^{\prime}$). By using a finite-temperature Lanczos method (FTLM), based on exact diagonalization of finite clusters, with $J_c$, $\phi$ and the $g$-factor as free parameters, the magnetic coupling constants $J$ and $J^{\prime}$ could be derived for Cs$_2$CuCl$_4$ and Cs$_2$CuBr$_4$ from fits to their $\chi_{mol}$($T$) data. As shown in Ref.\,\cite{Schmidt2015}, the so-derived values for $J$ and $J^{\prime}$ for \textit{x} = 0 and 4 are in excellent agreement with the corresponding values determined from high-field inelastic neutron scattering \cite{Coldea2002} and ESR measurements \cite{Zvyagin2014} with a reasonable $g$-factor expected for Cu$^{2+}$ in a tetrahedral environment \cite{Parker1971}. In view of the successful description of the border compounds \cite{Schmidt2015}, we apply the FTLM results to determine the magnetic coupling constants also for the stoichiometric mixed systems Cs$_2$CuCl$_3$Br$_1$ and Cs$_2$CuCl$_2$Br$_2$ by fitting their susceptibility data, see broken lines in Fig.\,1. The results of these fitting procedures, the quality of which for Cs$_2$CuCl$_3$Br$_1$ is about the same as for the border compounds, are also included in table I. It should be noted, that the compounds on the bromine rich side of the phase diagram, especially around the concentration x = 2, exhibit an upturn in the susceptibility \cite{Cong2011} at low temperature. As already addressed above with regard to the \textit{x} = 2 compound, this feature is an intrinsic property of unknown origin and not covered by the $J$-$J^{\prime}$ model. As a consequence, it is difficult to pinpoint the $J^{\prime}/J$ value for this compound by fitting the susceptibility because fits of almost identical quality could be obtained for a rather wide range of $J^{\prime}/J$ values ranging from 0.63 - 0.78 \cite{Tutsch2018a}.

In Ref. \cite{Tutsch2018} an alternative approach to determine $J^{\prime}/J$ has been discussed based on the analysis of low-temperature specific heat data. By applying a recently proposed spin Hartree-Fock description \cite{Werth2018} to the low-$T$ specific heat data, it was found that for the \textit{x} = 0, 1, and 4 compounds the so-derived $J^{\prime}/J$ values are very close to the ones obtained here by fitting the susceptibility data. However, for the \textit{x} = 2 compound, the spin Hartree-Fock results yield $J^{\prime}/J$ = 0.78 $\pm \,^{0.09}_{0.03}$ which lies at the upper end of the range $J^{\prime}/J \simeq$ 0.63 - 0.78 \cite{Tutsch2018a} obtained from fits to the susceptibility. It has been argued in Ref. \cite{Tutsch2018} that due to considerable uncertainties in the analysis of the susceptibility for this compound, the $J^{\prime}/J$ value derived from the specific heat is more reliable. We stress, as only relative changes of $J^{\prime}/J$ under pressure are of interest here, this uncertainty in $J^{\prime}/J$ for \textit{x} = 2 will not affect the conclusions drawn.

\section{Magnetic parameters}
As shown in the first column of table 1, the global energy scale \textit{J$_c$} in the Cs$_2$CuCl$_{4-x}$Br$_x$ series progressively grows with increasing Br content. On the other hand, the anisotropy parameter $\phi$ (second row in table 1) varies in a non-monotonous manner with the Br concentration. We assign this effect to the site-selective occupancy of the halide sublattice which mediates the magnetic coupling between the Cu$^{2+}$ $S$ = 1/2 moments. It was argued in Ref.\,\cite {Cong2011} that based on this structural peculiarity of Cs$_2$CuCl$_{4-x}$Br$_x$, one expects a significant increase of the degree of frustration $J^{\prime}/J$ on going from Cs$_2$CuCl$_4$ over Cs$_2$CuCl$_3$Br$_1$ to Cs$_2$CuCl$_2$Br$_2$. In fact, this trend is clearly reflected in the $J^{\prime}/J$ values derived from the above fitting procedure, see the right column of table 1, where $J^{\prime}/J$ increases from 0.33 (Cs$_2$CuCl$_4$) over 0.47 (Cs$_2$CuCl$_3$Br$_1$) to 0.63 - 0.78 \cite{Tutsch2018a} (Cs$_2$CuCl$_2$Br$_2$). We note that with $J^{\prime}/J \simeq$ 0.63 - 0.78 Cs$_2$CuCl$_2$Br$_2$ can be classified as a highly-frustrated triangular system.\\

\begin{figure}
\includegraphics[width=0.95\columnwidth]{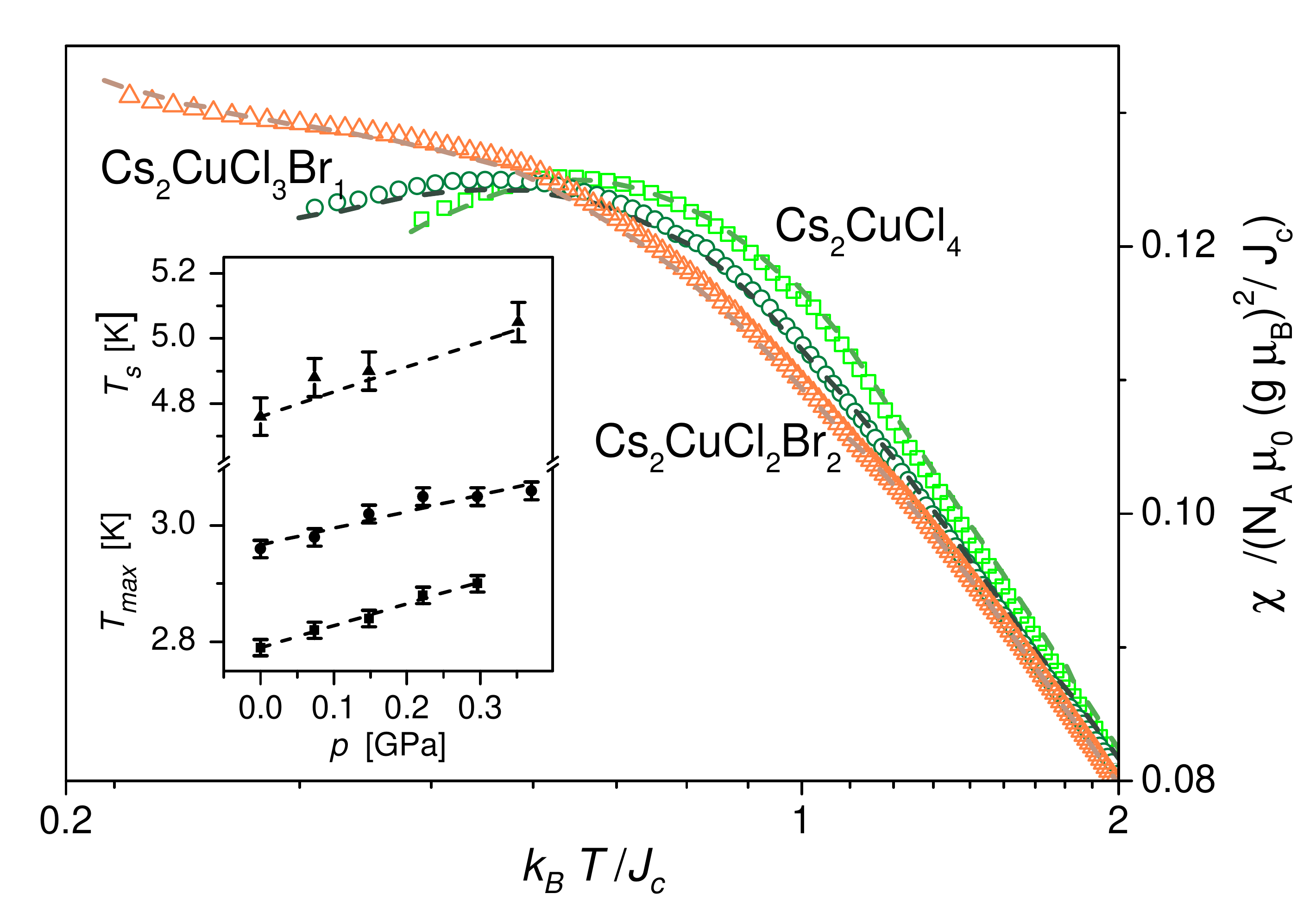}
\caption{(\emph{Color online}) $\chi_{mol}$($T$) of Cs$_2$CuCl$_4$ for $p$ = 0 (light green open squares) and $p_{max}$ = 0.296\,GPa (broken green line), $\chi_{mol}$($T$) for Cs$_2$CuCl$_3$Br$_1$ for $p$ = 0 (dark green open circles) and $p_{max}$ = 0.370\,GPa (broken black line) together with $\chi_{mol}$($T$) for Cs$_2$CuCl$_2$Br$_2$ for $p$ = 0 (orange open triangles) and $p_{max}$ = 0.352\,GPa (broken dark orange line) on a universal temperature and susceptibility scale. The inset shows the pressure dependence of $T_{max}$ for Cs$_2$CuCl$_4$ (black squares) for Cs$_2$CuCl$_3$Br$_1$ (black circles) and $T_{s}$ for Cs$_2$CuCl$_2$Br$_2$ (black triangles) together with a linear fit to the experimental data (broken black lines).}
\label{fig:3}
\end{figure}

After having determined the relevant magnetic parameters for the compounds under investigation at ambient pressure, we proceed by analyzing the effect of pressure on the magnetic susceptibility. Since there is a monotonous change of $\chi_{mol}$($T, p$) with increasing pressure in the pressure range investigated, we use the data taken at the maximum available pressure, $p_{max}$, where the effect is largest. To this end, we present in Fig.\,\ref{fig:3} the $\chi_{mol}$($T$) (symbols) and $\chi_{mol}$($T, p = p_{max}$) (lines) data for Cs$_2$CuCl$_4$, Cs$_2$CuCl$_3$Br$_1$ and Cs$_2$CuCl$_2$Br$_2$. The data are plotted in reduced units, i.e., normalized to the overall energy scale $J_c$ and by using dimensionless units for the ordinate. To account for the pressure dependence of $J_c$ = $J_c$($p$), we proceed as follows: according to Ref.\,\cite{Schmidt2015}, the position of the maximum in the susceptibility, $T_{max}$, is determined by $J_c$. Hence, by assuming $J_c \propto T_{max}$, we obtain $J_c^{-1}\cdot $d$J_c/$d$p$ = $T_{max}^{-1}\cdot $d$T_{max}/$d$p$ and calculate $J_c$($p_{max}$) from the experimentally-determined d$T_{max}/$d$p$ values. The so-derived $J_c$($p_{max}$) values are given in the left column of table 2.

Figure \ref{fig:3} discloses two main aspects. First, we find that the susceptibilities, plotted on a temperature scale normalized to $J_c$, for all three compounds remain practically unaffected by the application of pressure up to the maximum available pressure $p_{max}$ of 0.3 - 0.4\,GPa. This means that pressure in this range essentially affects $J_c$, while leaving the anisotropy ratio $J^{\prime}/J$ unchanged, implying that $J$ and $J^{\prime}$ are modified by pressure in the same way. Note that the collapse of the data at $p_{max}$ with the ambient-pressure data requires the usage of a weakly pressure-dependent $g$-factor, see table 2 for the used $g$ factor at $p_{max}$. We think that the slight reduction of $g$($p_{max}$) (by about 0.5\,\%) compared to the $g$ value at ambient pressure (table 1), reflects a small pressure-induced change in the local Cu environment which will be discussed in detail in the following section.

Second, despite using reduced units, the susceptibilities for the three compounds remain disjunct. For Cs$_2$CuCl$_4$ and Cs$_2$CuCl$_3$Br$_1$ the main difference lies in the position of the maxima rather than their heights, note the ordinate scale used in Fig.\,\ref{fig:3}, where the zero point is strongly suppressed. It was found in Ref.\,\cite{Schmidt2015} that $\chi_{mol}^{max}$ varies only slightly over a wide range of frustration parameters, with the normalized $\chi_{mol}^{max}$ staying constant at around $\sim$ 0.126 for $J^{\prime}/J$ varying between 1 and 0.32. On the other hand, for the same range of frustration parameter, $T_{max}$ changes considerably from 0.35$J_c$ to 0.6$J_c$. Hence, the separation of the normalized susceptibility curves in Fig.\,\ref{fig:3} reflects the different degree of frustration in these two compounds. Even though the susceptibility for Cs$_2$CuCl$_2$Br$_2$ is somewhat different in lacking a clear maximum, it follows the same trend. We therefore conclude that the effect of pressure up to about 0.4\,GPa, the maximum pressure used here, applied to Cs$_2$CuCl$_4$, Cs$_2$CuCl$_3$Br$_1$ and Cs$_2$CuCl$_2$Br$_2$ is to modify the systems' overall energy scale $J_c$ while leaving their degree of frustration unaffected. The frustration does change, however, by substituting Cl by Br.

\section{Magnetoelastic coupling constants}
 From the pressure dependence of $T_{max}$, respectively $T_{s}$, for the various compounds, and the corresponding d$J_c$/d$p$ values, we can determine the magnetoelastic coupling constants defined as d$J_c$/d$\epsilon_V$ = $K_0\cdot$ d$J_c$/d$p$, where $\epsilon_V$ = $\Delta V/V$ denotes the volume strain and $K_0$ = -$V \cdot$ d$p$/d$V$ the bulk modulus. By using $K_0$ = (15.0 $\pm$ 0.2)\,GPa for Cs$_2$CuCl$_4$ \cite{Xu2000} as a reasonable approximation also for the Br-substituted compounds, we obtain $k_{B}^{-1}$d$J_c$/d$\epsilon_V$ values of (5.7 $\pm$ 0.4)\,K for Cs$_2$CuCl$_4$ and (4.4 $\pm$ 0.8)\,K for Cs$_2$CuCl$_3$Br$_1$. For Cs$_2$CuCl$_2$Br$_2$, where we take the pressure dependence of the shoulder in $\chi_{mol}$($T, p$) for the determination of d$J_c$/d$p$, we find $k_{B}^{-1}$d$J_c$/d$\epsilon_V$ = (11.6 $\pm$ 2.8)\,K. These magnetoelastic coupling constants, determined directly from measurements under hydrostatic pressure, are comparable to the values $k_{B}^{-1}$d$J$/d$\epsilon_L \sim$ 2.8\,K derived from ultrasonic measurements on Cs$_2$CuCl$_4$, where $\epsilon_L$ denotes the strain induced by the longitudinal modes c$_{11}$ and c$_{33}$ \cite{Cong2016}.  These values, all lying in the range of a few degrees Kelvin, are remarkably small -- at least two orders of magnitude smaller compared to other low-dimensional Cu-based quantum magnets.\cite {Wolf2001,Cong2014} As will be discussed below in more detail, we assign the small magnetoelastic coupling here to a structural peculiarity of the present system.

\begin{figure}
\includegraphics[width=0.95\columnwidth]{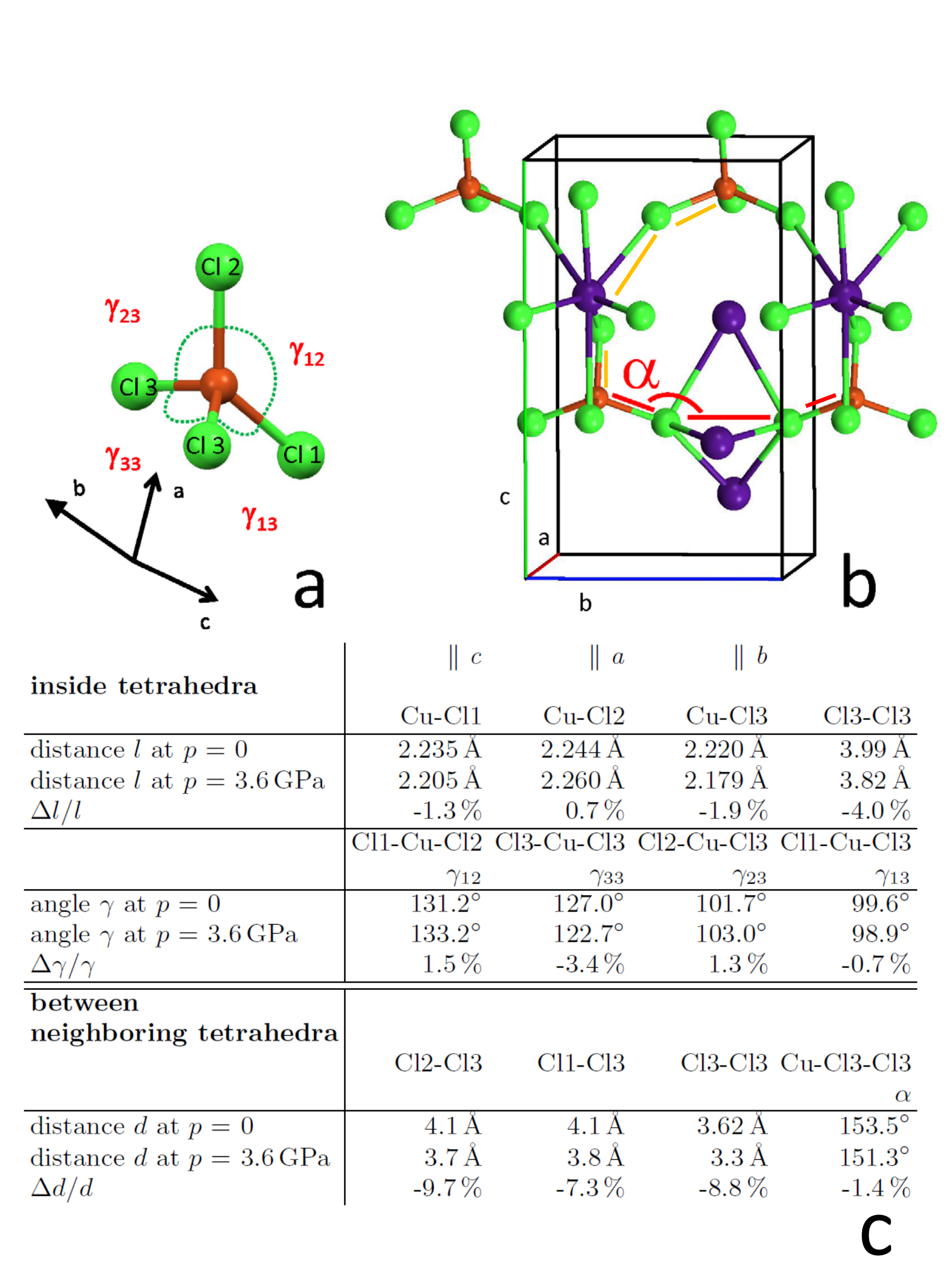}
\caption{(\emph{Color online}) \textbf{a}: Single Cu-halide tetrahedron in Cs$_2$CuCl$_{4}$ at ambient pressure according to the data reported in Ref.\,\cite{Xu2000}. Due to the Jahn-Teller effect of the Cu$^{2+}$ ion (red-brown spheres), the tetrahedron is distorted. As shown in table 3, the angles $\gamma_{12}$ and $\gamma_{33}$ are significantly enlarged whereas $\gamma_{13}$ and $\gamma_{23}$ are reduced with respect to the regular tetrahedron angle of 109.5$^\circ$. The influence of steric pressure due to the crystal structure leads to a further lowering of the symmetry of the tetrahedron visible in the difference between $\gamma_{12}$ and $\gamma_{33}$ or $\gamma_{13}$ and $\gamma_{23}$, respectively.
\textbf{b}: Array of Cu-halide tetrahedra in Cs$_2$CuCl$_{4}$  together with Cs ions (purple spheres) in the \textit{b, c} - plane. Note, that some of the halide atoms are omitted for clarity. The angle $\alpha$ is indicated which is relevant for the dominant magnetic coupling constant \textit{J} as shown in Ref. \cite{Roser2011}. The red line indicates the exchange path for $J$ and the orange line indicates the exchange path for $J^{\prime}$. \textbf{c}: Experimental atomic distances and binding angles of Cs$_2$CuCl$_4$ at ambient pressure and $p$ = 3.6\,GPa taken from Ref.\,\cite {Xu2000}. See main text for more details.}
\label{tetraeder}
\end{figure}

\begin{table*}[t]
  \caption{Atomic distances and binding angles of Cs$_2$CuCl$_4$ from structure optimization simulations at ambient pressure (Theoretical 1.0 GPa), 0.5 GPa (Theoretical 1.5 GPa) and 3.5 GPa (Theoretical 4.5 Gpa).}
\begin{tabular}{r|rrrr}
\hline
\textbf{inside tetrahedra} & Cu-Cl1 & Cu-Cl2 & Cu-Cl3 & Cl3-Cl3 \\
\hline
distance $l$ at $p$ = 0 & 2.263\,{\AA} & 2.271\,{\AA} & 2.238\,{\AA} & 4.034\,{\AA}    \\
$\Delta l/l$ (0.5 Gpa) & -0.2\,\% & -0.1\,\% & -0.3\,\% & -0.6\,\%    \\
$\Delta l/l$ (3.5 Gpa) & -1.5\,\% & -0.8\,\% & -1.5\,\% & -3.6\,\%    \\
\hline
 & Cl1-Cu-Cl2 & Cl3-Cu-Cl3 & Cl2-Cu1-Cl3 & Cl1-Cu-Cl3 \\
& $\gamma_{12}$ & $\gamma_{33}$ & $\gamma_{23}$& $\gamma_{13}$ \\
\hline
angle $\gamma$ at $p$ = 0 & 132.2 $^{\circ}$ & 128.6 $^{\circ}$  & 101.2 $^{\circ}$ & 99.06 $^{\circ}$   \\
$\Delta \gamma/\gamma$ (0.5 Gpa) & 0.1 \,\% & -0.5\,\% & 0.2 \,\% &  -0.01\,\%  \\
$\Delta \gamma/\gamma$ (3.5 Gpa) & 1.1 \,\% & -3.9 \,\% & 1.2\,\% &  -0.2\,\%  \\
\hline
\hline
\textbf{between neighboring tetrahedra} & Cl1-Cl3 & Cl3-Cl3 & & Cu1-Cl3-Cl3  \\
   &  &  &  &$\alpha$ \\
\hline
distance $d$ at $p$ = 0 & 4.106 \,{\AA} & 3.54 \,{\AA} & & 154.3 $^{\circ}$ \\
$\Delta d/d$ (0.5 Gpa) & -1.3 \,\% & -1.6 \,\% & & -0.2 \,\%  \\
$\Delta d/d$ (3.5 Gpa) & -6.3 \,\% & -7.6 \,\% & & -1.6 \,\%  \\
\end{tabular}
\end{table*}

\begin{table*}[ht]
\caption {Lattice constants, Cu-Cu distances in the \textit{b,c} plane $d_J$ and $d_{J^{\prime}}$ and calculated exchange interactions $J$, $J^{\prime}$ in K for Cs$_2$CuCl$_4$ at various pressures.}
\begin{tabular}{lrrr|rrr}
\hline
 &  lattice constant (a,b,c)  & $d_J$  & $d_{J^{\prime}}$ &\textit{J/$k_B$}\,(K) &$J^{\prime}$/$k_B$ \,(K)  & \textit{J}/$J^{\prime}$ \\
\hline
\textbf{Experimental structure} \cite{McGinnety1972,Xu2000} & & & & & & \\
{ ambient pressure } & (9.760, 7.609, 12.397)               &7.609\,{\AA}   & 7.283\,{\AA} &3.632    &2.135   &0.59  \\
{ 3.6 GPa }   &  (9.254, 7.142, 11.787)                  &7.142\,{\AA}   & 6.902\,{\AA} &4.885    &3.644   &0.75  \\
\textbf{Theoretical structure} & & & & & & \\
{ 1 GPa  }     &  (9.770, 7.574, 12.330)  & 7.574\,{\AA}  & 7.245\,{\AA} &5.257    &2.054   &0.39  \\
{ 1.5 GPa} &  (9.677, 7.495, 12.222)                    & 7.495\,{\AA}  & 7.179\,{\AA} &5.582    &2.274   &0.41  \\
{ 4.5 GPa }    &  (9.309, 7.160, 11.787)  & 7.160\,{\AA}  & 6.907\,{\AA} &6.649    &3.493   &0.53  \\
\end{tabular}

\end{table*}

\section{Magneto-structural correlations}
For the discussion of the magneto-structural correlations in the compounds of Cs$_2$CuCl$_{4-x}$Br$_x$ (0 $\leq$ \textit{x} $\leq$ 2) it is useful to recall the results of structural investigations under pressure, so far available only for the x = 0 border compound. According to these x-ray studies, performed up to 4.0\,GPa \cite{Xu2000}, the structure stays orthorhombic, space group \textit{Pnma}, up to 3.6\,GPa with reduced lattice parameters compared to those at ambient pressure \cite{Morosin1961,McGinnety1972,Bailleul1991}. At low pressures up to about 0.6\,GPa, which is relevant for the present study, the changes in the lattice parameters were found to be linear with pressure \cite{Xu2000}. At 0.6\,GPa, the relative reduction of the lattice parameters along the $a$- and $c$-axis are of similar size, amounting to about 0.9\,\%, which is a factor of 2 smaller than the compression of the $b$-axis. Such a volume compression of about 4\,\% at 0.6\,GPa, which increases to about 15.3\,\% at 3.6\,GPa \cite{Xu2000}, is rather large for an inorganic crystalline material, but is typical for Cs salts. \\

The x-ray study in Ref.\,\cite {Xu2000} revealed some structural peculiarities of Cs$_2$CuCl$_{4}$ which are summarized in the table shown in Fig. 6c. Here we display the structural parameters inside the distorted Cu-halide tetrahedron (see Fig.\,\ref{tetraeder}a for the labelling of the atoms and angles) and those between neighboring tetrahedra. We compare these parameter at ambient pressure with those at 3.6\,GPa. As discussed in Refs.\,\cite{Morosin1961,Roser2011}, the distortion of the tetrahedron from a regular shape (with all angles being 109.5$^{\circ}$) is mainly due to the Jahn-Teller effect of the Cu$^{2+}$ ion, lifting the degeneracy of the Cu t$_{2g}$ orbitals, and a smaller effect resulting from the steric pressure exerted by the Cs ions. The latter effect is responsible for the asymmetric changes in the angles $\gamma_{12}$ vs $\gamma_{33}$ and $\gamma_{13}$ vs $\gamma_{23}$, cf.\,table 3, see also Ref.\,\cite{Roser2011}.
The table in Fig. 6c also includes the pressure-induced reduction of the bond lengths, $\Delta l/ l$, and angles, $\Delta \gamma/ \gamma$ inside the tetrahedra between Cu and the four halide atoms (Cl1, Cl2, and Cl3 (2$\times)$) at $p$ = 3.6\,GPa. All changes are small and slightly anisotropic. They amount to about $\pm$1\,\% for the bonds along the $a$- and $c$-axis, around 2\,\% for the chain direction ($b$-axis) and up to -3.4\,\% for the tetrahedron angle $\gamma_{33}$ at $p$ = 3.6\,GPa. This angle is closely related to the angle $\alpha$ characterising the Cu-Cl3-Cl3-Cu exchange path, see Fig. 6b. As shown in Fig. 6c, hydrostatic pressure of 3.6 GPa slightly enhances the steric effect and changes the Jahn-Teller distortion only to a small extent. In fact, as shown by these structural investigations, the local Cu environment is only weakly modified by hydrostatic pressure. This provides a natural explanation for the tiny $p$-induced changes in the $g$-factor revealed from the fits to the susceptibility under pressure, cf.\,table 2. It was pointed out in Ref. \cite{Roser2011} that the Cu-Cl3-Cl3-Cu exchange path which is influenced by $\alpha$ and the distance of the chlorine atoms along the \textit{b} direction is responsible for the dominant magnetic coupling constant \textit{J}. In Fig. 6b it is shown that the Cs ions (purple spheres) are in a nearly spherical environment of the halide atoms belonging to different Cu-halide tetrahedra. Salts containing Cs ions generally exhibit a large compressibility and in combination with this spherical environment this leads to large nearly isotropic length changes between the Cu-halide tetrahedra as displayed in table 3. The changes of the halide-halide distances between neighboring Cu-halide tetrahedra from 7.3 \% to  9.7 \% at 3.6 GPa are extraordinarily large. Assuming that similar to the Cu-Cl3-Cl3-Cu exchange path which is responsible for $J$, the path Cu-Cl3-Cl1-Cu mediates $J^{\prime}$, it seems reasonable that both coupling constants change in a similar way under pressure. In contrast to these large pressure-induced changes of the halide distances between the tetrahedra, $\alpha$ is only reduced to about -1.4 \% at 3.6 GPa. Such a reduction of $\alpha$ with pressure should lead to a reduction of $J$. However, this effect is overcompensated by the increased overlap of the halide orbitals along the Cu-Cl3-Cl3-Cu exchange path.

In order to rationalize these experimental observations, we performed {\it ab initio} density functional theory calculations for Cs$_2$CuCl$_{4}$ under pressure and obtained the exchange coupling constants at various pressure values.

\section{Theory details}
For our microscopic analysis of the origin of the non-monotonous evolution of the exchange coupling constants as a function of the bromine concentration x extracted from experiment,  structure optimization simulations under pressure were performed with the projector-augmented wave method~\cite{Bloechl1994,Kresse1999} as implemented in the VASP code~\cite{Kresse1993,Kresse1996,Kresse1996b}. We used the generalized gradient approximation (GGA)~\cite{Perdew1996} as exchange-correlation functional and included correlation corrections within the GGA+U scheme~\cite{Liechtenstein1995} with  effective correlation strength parameters $U$ = 6 eV and $J_{\rm H}$ = 1 eV (Hund's coupling).
Relaxations were performed assuming a ferromagnetic order of the Cu atoms. Convergence of the properties of interest was achieved for a 10 $\times$ 10 $\times$ 10 \textit{k}-mesh and an energy cutoff of 600 eV. We observe that the structural data of the experimental structure at ambient pressure (0 GPa) (see Table 3) are rather well reproduced by the theoretically relaxed structure at 1 GPa. This provides the reference for comparison between theory and experiment. Likewise, the experimental crystal structure at 3.6 GPa (see table 4) compares with the theoretical structure at 4.5 GPa and we expect the theoretical simulations at 1.5 GPa to provide a good reference for the measurements performed at 0.5 GPa. Given that the experimental x-ray data were obtained at finite temperature and our simulations are at zero temperature with a certain choice of exchange correlation functional and magnetism, the agreement is very good.

We calculated the anisotropic triangular-lattice exchange parameters $J$ and $J^{\prime}$ from total energy calculations of various supercells~\cite{Roser2011,Tutsch2014} for both, the experimental structures at hand, as well as the simulated theoretical structures under pressure. The total energy calculations were performed on the basis of the full-potential non-orthogonal local-orbital basis (FPLO)~\cite{Koepernik1999}, employing the generalized gradient approximation (GGA)~\cite{Perdew1996} as well as the around mean field (AMF) version of GGA+U~\cite{Czyzyk1994} with $U$ = 6 eV and $J_{\rm H}$ = 1 eV. We chose a 5 $\times$ 4 $\times$ 3 mesh of $k$ points for the supercell Brillouin zone integration.

Table 4 shows the calculated exchange interactions. The different structural details influence the electronic properties and lead to different $J^{\prime}$/$J$ values. Both sets of structures, the experimental 0 GPa and 3.6 GPa as well as the theoretical 1 GPa and 4.5 GPa structures show the same trend, i.e. that $J^{\prime}$/$J$ increases with increasing pressure
with similar ratio values. This gives us further assurance to select the theoretical 1.5 GPa in order to discuss the 0.5 GPa measurements. Interestingly, we observe that the ratio $J^{\prime}$/$J$ remains almost unchanged upon the application of weak pressures, in agreement with the experimental susceptibility measurements. A closer look at the structural and electronic structure reveals that for these pressures there are no significant changes. Only when high pressures are applied, does the system show an increased $J^{\prime}$/$J$ ratio due to larger modifications that affect  $J^{\prime}$ more strongly than $J$.

\section{Conclusion and Summary}
From the magnetic susceptibility measurements under variable hydrostatic He-gas pressure performed on various single crystals of Cs$_2$CuCl$_{4-x}$Br$_x$ with x = 0, 1, and 2 we extract the magnetic exchange coupling constants $J$ and $J^{\prime}$ based on the recently proposed $J$-$J^{\prime}$ model \cite{Schmidt2015}. This enables us to determine the degree of frustration $J^{\prime}$/$J$ for these stoichiometric compounds. With $J^{\prime}$/$J$ $\simeq$ 0.63 - 0.78 for Cs$_2$CuCl$_2$Br$_2$ this system exhibits the highest degree of frustration followed by $J^{\prime}$/$J$ = 0.47 for Cs$_2$CuCl$_3$Br$_1$ and $J^{\prime}$/$J$ = 0.40 for Cs$_2$CuBr$_4$. The pure chlorine compound, x = 0, has the smallest degree of frustration with $J^{\prime}$/$J$ = 0.32. From the pressure dependence of the magnetic susceptibility up to about 0.4\,GPa we find that the overall energy scale, given by $J_c = (J^2$ + $J^{\prime 2}$)$^{1/2}$, increases under pressure, whereas the degree of frustration $J^{\prime}$/$J$ remains unchanged in this pressure range. For a microscopic analysis of the magnetic exchange coupling constants and their pressure dependences we perform structure optimization simulations. The structural data of the experimental structure at ambient pressure (0 GPa) are rather well reproduced by the theoretically relaxed structure at 1 GPa. We observe in this calculations that the ratio $J^{\prime}$/$J$ remains almost unchanged upon the application of small pressures. This result is in agreement with the experimental findings. Only when higher pressures are applied, the systems show an increased $J^{\prime}$/$J$ ratio due to larger modifications that affect  $J^{\prime}$ more strongly than $J$. For the magnetoelastic coupling constants, extraordinarily small values are obtained. They are about two orders of magnitude smaller compared to other Cu-based quantum magnets. We assign these observations to a structural peculiarity of this class of materials, consisting of well-isolated magnetic units not sharing any common coordination element.

\section{acknowledgement}
Work done at Goethe University Frankfurt was supported by the German Science Foundation (DFG) through the SFB/TR\,49. We thank N. van Well for her help in the sample preparation.

\end{document}